\documentclass{aastex63} 

\accepted{July 21, 2021}

\usepackage{apjfonts}

\def\lae{\mathrel{\raise .4ex\hbox{\rlap{$<$}\lower 1.2ex\hbox{$\sim$}}}}
\def\gae{\mathrel{\raise .4ex\hbox{\rlap{$>$}\lower 1.2ex\hbox{$\sim$}}}}

\usepackage{amsmath}
\usepackage{comment}
\usepackage{appendix}
\shorttitle{Polarimetry Data with Angular Uncertainties}
\shortauthors{Marshall}

\begin{document}

\title{Analysis of Polarimetry Data with Angular Uncertainties}

\correspondingauthor{Herman L.\ Marshall}
\email{hermanm@space.mit.edu}

\author{Herman L.\ Marshall}
\affiliation{Kavli Institute for Astrophysics and Space Research,\\
 Massachusetts Institute of Technology, 77 Massachusetts Ave.,
 Cambridge, MA 02139, USA}
 
\begin{abstract}

For a track based polarimeter, such as the Imaging X-ray Polarimetry Explorer, (IXPE),
the sensitivity to polarization depends on the modulation factor, which is a strong
function of energy.
In previous work, a likelihood method was developed that would account for
this variation in order to estimate the minimum detectable polarization (MDP).
That method essentially required that the position angles of individual events should be
known precisely.
In a separate work, however, it was shown that using a machine learning method for
measuring event tracks can generate track angle uncertainties, which can be used
in the analysis.
Here, the maximum likelihood method is used as a basis for revising the estimate
of the MDP in a general way that can include uncertainties in event
track position angles.  The resultant MDP depends solely upon the distribution
of track angle uncertainties present in the input data.  Due to the physics of the IXPE detectors, it is possible to
derive a simple relationship between these angular uncertainties and the
energy-dependent modulation
function as a step in the process.

\end{abstract}


\section{Introduction}

X-ray polarimeters that use the photoelectric effect,
such as the gas pixel detectors (GPDs) \citep{2001Natur.411..662C,2007NIMPA.579..853B,2012AdSpR..49..143F,2021arXiv210705496B}
in the Imaging X-ray Polarimetry Explorer \citep[IXPE:][]{2016SPIE.9905E..17W,ixpe}
provide a signal that is probabilistically related to the nput polarization direction
A histogram of event phase angles, $\psi$, then has a characteristic
instrument-dependent
modulation factor, $\mu$, defined as the amplitude of the signal modulation
for a source of 100\% linearly polarized photons.
Instrument models and ground-based calibration using sources of known polarization angle are used to
determine how $\mu$ depends on energy $E$.
Generally, polarimeters are characterized by their minimum detectable polarization (MDP) at a given level of confidence
(e.g., 99\%) which depends on $\mu$:

\begin{equation}
{\rm MDP}_{99} = \frac{4.292}{\mu R_S} [\frac{R_S + R_B}{T}]^{1/2}
\end{equation}

\noindent
where $T$ is the exposure time and
the count rates the instrument records for both the source $R_S$ and background $R_B$
\citep{weisskopf:77320E}.
When background is negligible, then

\begin{equation}
\label{eq:mdp}
{\rm MDP}_{99} = \frac{4.292}{\mu N^{1/2}}
\end{equation}

\noindent
where $N = T R_S$ is the number of counts in the observation.
In a previous paper, we showed how to modify Eq.~\ref{eq:mdp} for the 
case where $\mu$ is a strong function of $E$ \cite[][hereafter Paper I]{marshall20}.
In this paper, we extend the method to include uncertainties in the event phase angles.

Estimating the uncertainties in the phase angles was a major feature
of a new advance in handling event tracks based on convolutional neural networks \cite[CNNs,][]{PEIRSON2021164740}.
\citet{PEIRSON2021164740} showed that a  machine learning approach could provide
significantly larger modulation factors than previous empirical methods, yielding
improvements to the MDP for a given simulated observation.
In order to choose networks yielding the smallest MDPs, a weighting term
of form $\sigma^{-\lambda}$ was applied to the likelihood
based on the event uncertainties, $\sigma$.
However, the weighting term was heuristic, with a free parameter, $\lambda$, that was
set {\it a posteriori} using simulations in order to provide small MDPs.  Thus,
the eventual MDP values actually depended on the free parameter.
\citet{PEIRSON2021164740} found that $\lambda$ should be 1.5-2 but that the
value of $\lambda$ that gives the minimum MDP depends
on the spectral shape of the source, which determines the number of events
with low energy and high $\sigma$ relative to the number with high energy and
low $\sigma$.  The choice of an optimal CNN then depends on the spectral shape of the source
and $\lambda$.

Here, I demonstrate that
these track angle uncertainties can be directly related to the modulation function.
Furthermore, I
show how a maximum likelihood analysis of an observed data set can be modified
to include the event angle uncertainty information in an objective manner.
Using this approach, the MDP is derived for the case of large $N$ and is shown
to depend on the distribution of track angle uncertainties present in the data.
This distribution necessarily depends on source spectrum, so the MDP
will also depend on the source spectrum.
However, the process for determining the
optimal CNN of a set of possibilities will depend on its accuracy in measuring track
angles and not directly on the source spectrum and will not require a weighting term
that depends on a free parameter.

\section{Background}

We follow the unbinned method of our previous paper, \cite[][hereafter Paper I]{marshall20} in formulating
the problem and defining quantities.
An observation with the instrument consists of $N$ events with index $i$, where each event is
tagged by its energy, $E_i$, and instrument phase, $\psi_i$.
In the original formulation, it was assumed that this phase is the perfectly measured position angle
of the event track.
At energy $E$, the instrument response is characterized by a modulation factor is $\mu_E$ and
the instrument effective area is $A_E = A \mathcal{A}(E)$, where $A$ is a constant with
units of cm$^{-2}$.
For reasons that will be clear later, the intrinsic source flux is defined
to be $n_E = n_0 \mathcal{N}(E_i)$,
where the function $\mathcal{N}(E)$ is a function giving the spectral shape of the source, and $n_0$ is
the normalization of the spectrum in suitable units (say ph/cm$^2$/s/keV per radian of rotation).
In this model, the event density in a differential energy-phase element $dE d\psi$ about $(E, \psi)$ is

\begin{equation}
\lambda(E, \psi) = [ 1 +  \mu_E  (q \cos 2\psi + u \sin 2\psi) ] n_0 A \mathcal{N}(E) \mathcal{A}(E) T dE d\psi
\end{equation}

\noindent
where $T$ is the exposure time, and $q$ and $u$ are the fractional Stokes parameters, given by
$Q = P \cos \phi_0 = q I$ and
$U = P \sin \phi_0 = u I$, respectively.  The (linear) polarization fraction
is then $p \equiv P/I = (q^2 + u^2)^{1/2}$ and the source phase angle is $\phi_0 = \tan^{-1} u/q$.
The electric vector position angle (EVPA) is $\varphi = \phi_0/2$.

In general, the normalized Stokes parameters $q$ and $u$ are functions of energy and a forward
folding approach (such as {\tt xspec}) is needed to estimate parameters of those functions.
For now, I assume that $q$ and $u$
are independent of $E$ or that the bandpass of interest is sufficiently narrow that the
polarization fraction and EVPA can be assumed to be constant across the band.
For some sources, such an assumption may be required in order to detect
a weak polarization signal or to place a limit on the degree of polarization.

The log-likelihood for a Poisson probability distribution of events, $S = -2 \ln \mathcal{L} $, is

\begin{eqnarray}
S(n_0, q, u) & = & -2 \sum_i \ln \lambda(E_i, \psi_i) +
2 T \int n_E A_E dE  \int_0^{2\pi} [ 1 +  \mu(E)  (q \cos 2\psi + u \sin 2\psi) ]  d\psi \\
 & = & -2 \sum_i  \ln ( 1 +  q \mu_i \cos 2\psi_i + u \mu_i \sin 2\psi_i) - 2 N \ln(n_0) +
  4 \pi T n_0 A \int \mathcal{N}(E) \mathcal{A}(E) dE
\label{eq:unbin}
\end{eqnarray}

\noindent
where $\mu_i = \mu(E_i)$.  The best estimate of $n_0$ is independent of $q$ and $u$:
\begin{equation}
\hat{n_0} = \frac{N}{2 \pi A T \int \mathcal{N}(E) \mathcal{A}(E) dE} ~~~~.
\label{eq:nzero}
\end{equation}

\noindent
The log-likelihood for the polarization parameters alone is merely the first term of Eq.~\ref{eq:unbin}:

\begin{equation}
\tilde{S}(q,u) =  -2 \sum_i  \ln ( 1 +  q \mu_i \cos 2\psi_i + u \mu_i \sin 2\psi_i)   ,
\label{eq:like}
\end{equation}

\noindent
and the MDP is

\begin{equation}
{\rm MDP}_{99}  \approx \frac{4.292}{(\bar{\mu^2} N)^{1/2}}
\end{equation}

\noindent
when $N$ is large, where $\bar{\mu^2}$ is the
count-weighted average of $\mu^2$ (Paper I).

\section{Deriving the Modulation Factor}

For a photoelectron polarimeter, the interaction cross section is proportional to $\sin^2 \theta \cos^2 \psi$, where
$\theta$ is the polar angle and $\psi$ is the azimuthal angle relative to the input photon's $E$ vector, $\varphi$.
For IXPE, the charge generated by photoelectron interactions in the detector gas
drifts in an applied electric field in before collection at the pixelized anode \cite[cf.][]{2007NIMPA.579..853B}.
Thus, the polar angle is not measured and the azimuthal angle is independent
of energy, so the probability distribution of track angles is $p(\psi) = 2\cos^2 \psi-\varphi
= 1 + \cos 2 (\psi-\varphi)$ for 100\% polarized X-rays ($p$ = 1); thus, $\mu = 1$ in theory.
Noting that this distribution is only satisfied for a
perfect polarimeter, we examine the hypothesis that the modulation factor is less than 1 due to track angle measurement
uncertainties.  A similar hypothesis was examined by \citet{2018Galax...6...46V}.  They took an empirical approach
and used Monte Carlo methods to confirm the relationship between track measurement uncertainties and the modulation
function.  However, now that a track measurement method has been developed that
provides uncertainties for individual tracks \citep{PEIRSON2021164740}, we explore an analytical approach to the modulation function.

We start by assuming that the angle error distribution can be modeled by a single function, $G(\psi; \psi', \alpha)$, where
$\psi$ is the track's measured phase angle, $\psi'$ is the true phase angle, and $\alpha$ is a
generic parameter that characterizes the error
distribution.  For this study, we consider two functional forms of $G$: a Gaussian, with $\alpha = \sigma$ as the
standard deviation of the Gaussian, and a von Mises distribution, with $\alpha = \kappa$.  The von Mises distribution
is circular, which may be more appropriate for azimuthal angles.
The forms of the two distributions are

\begin{eqnarray}
G(\psi; \psi', \sigma) & = & \frac{1}{\sqrt{2 \pi} \sigma} e^{-\frac{(\psi - \psi')^2}{2\sigma^2}} \\
G(\psi; \psi', \kappa) & = & \frac{1}{2 \pi I_0(\kappa)} e^{\kappa \cos(\psi - \psi')}
\end{eqnarray}

\noindent
where $I_0(\kappa)$ is the modified Bessel function of order 0.  Note that the two distributions are practically
equivalent for small $\sigma$ and large $\kappa$, with $\kappa \sim \sigma^{-2}$.

\begin{figure}
    \centering
    \includegraphics[width=\textwidth]{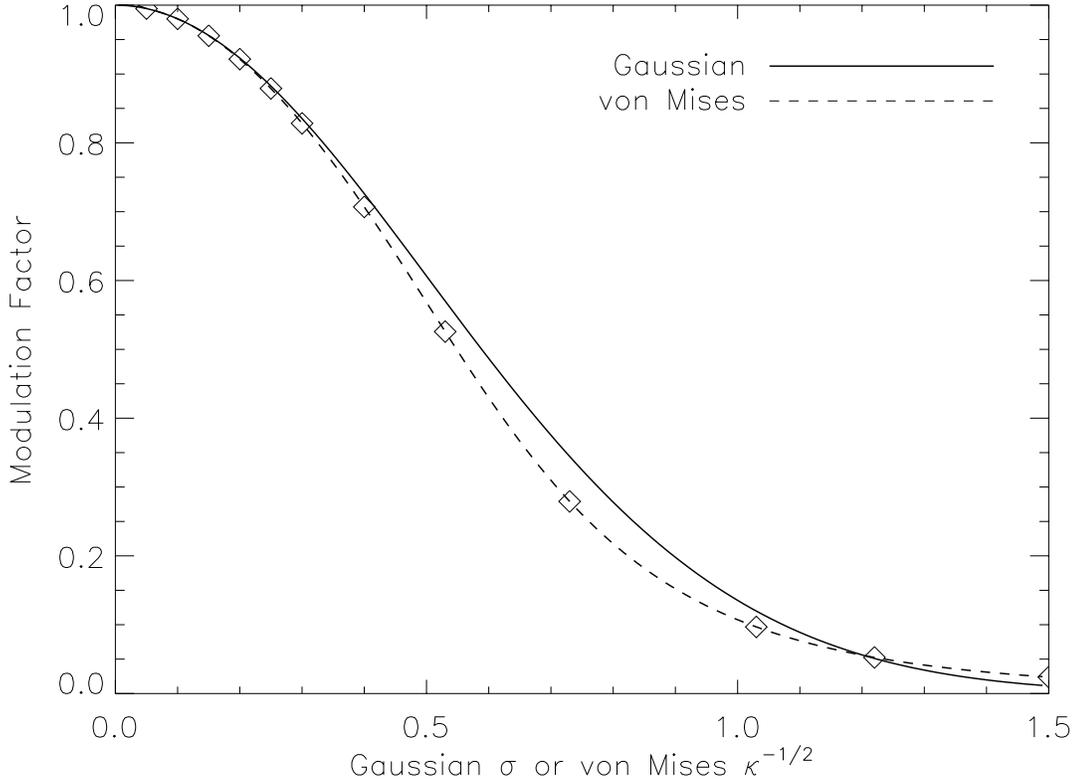}
    \caption{Comparison of the modulation factor from a Gaussian model
    with standard deviation $\sigma$ to that
    using a von Mises distribution with parameter $\kappa$.
   }
    \label{fig:vonmises}
\end{figure}

We can now compute the probability distribution for the observed track angles by integrating
over the distribution of true (unknown) angles:

\begin{equation}
p(\psi) = 1+ \int G(\psi; \psi',\alpha) \cos 2(\psi' - \varphi ) ~d\psi' = 1 + \mu'(\alpha) \cos 2(\psi - \varphi )
\label{eq:modcounts}
\end{equation}

\noindent
where we find the expected modulation factors depend solely on the characteristic parameters of
the input error distributions:

\begin{eqnarray}
\label{eq:gaussian}
\mu'(\sigma) & = & e^{-2\sigma^2} \\
\mu'(\kappa) & = & \frac{I_2(\kappa)}{I_0(\kappa)}
\label{eq:vonmises}
\end{eqnarray}

\noindent
for the Gaussian and von Mises distributions, respectively.  In Fig.~\ref{fig:vonmises}, we compare these two distributions,
noting that the two are practically identical for small (Gaussian) uncertainties, as expected, and that the general forms
are quite similar.

Before exercising this development, however, we must recognize that $\mu$ is usually given as a function of energy
while the distributions of uncertainties at any given energy are not delta functions and are not even unimodal at
some energies.  See Fig. 4 of \citet{PEIRSON2021164740} for two examples.  While the distribution of $\sigma$ is
approximately Gaussian with a mean of $\sigma = 0.62$ rad at 3 keV, the 6.4 keV distribution shows a peak at $\sigma = 0.33$ rad
and a broad, weaker peak at 0.6 rad.  If we designate the distribution of $\sigma$ for a given energy by $\rho(\sigma; E)$,
normalized so that $\int \rho(\sigma; E) d\sigma = 1$, we can
then estimate the observed modulation function by

\begin{equation}
\mu(E) = \int \rho(\sigma;E) \mu'(\sigma) d\sigma =  \int \rho(\sigma;E) e^{-2\sigma^2} d\sigma
\label{eq:mu}
\end{equation}

\noindent
for the Gaussian case.  The von Mises case is analogously determined.

For 3 keV, Eq.~\ref{eq:gaussian} gives $\mu' = 0.46$.
Approximating the $\rho(\sigma)$ distribution at 3 keV
as a single Gaussian with mean $\sigma_0$ and standard deviation $s_{\sigma}$, then

\begin{equation}
\mu = \frac{1}{\sqrt{4 s_{\sigma}^2+1}} e^{ \frac{-2\sigma^2}{4 s_{\sigma}^2+1} }
\end{equation}

\noindent
For $E = 3$ keV, $\sigma_0 = 0.62$ and $s_{\sigma} = 0.076$ and $\mu$ increases, but only by 0.5\%
compared to assuming that $\rho(\sigma)$ is a delta function at $\sigma_0$.
Because $s_{\sigma}^2 \ll \sigma_0^2$, the result is not sensitive
to $s_{\sigma}$, changing by less than 0.3\% for 10\% changes in $s_{\sigma}$.
The best value of $\mu$ using CNNs was
just under 0.40, with other methods giving $\mu$ as low as $0.32$ \citep{PEIRSON2021164740}.
While a few percent loss may be expected due to events that interact in the detector window or the gas
electron multiplier (GEM), this effect is insufficient to explain the difference between the model
and the CNN results and these events are typically cut out of the analysis anyway.
We can recover the observed modulation factor by adding a systemic variance term, replacing
$\sigma_0^2$ with $\sigma_0^2 + \sigma_s^2$; for $\sigma_s = 0.25$, the predicted value of $\mu$
drops to 0.41.
This systemic variance is not to be identified with the ``epistemic'' variance term added by \cite{PEIRSON2021164740}
to compute total track angle variance.
The predicted $\mu$ based on the von Mises error distribution may prove
to be better, requiring knowledge of the $\kappa$ distribution characterizing the track measurements
which is outside the scope of this paper.

At 6.4 keV, we approximate the $\rho(\sigma)$ function as the sum of two Gaussians, one with
70\% of the events at $\sigma_0 = 0.33$ and $s_{\sigma} = 0.076$ and 30\% with $\sigma_0 = 0.58$
and $s_{\sigma} = 0.076$.
With these values,
$\mu = 0.72$, which compares well to the values obtained from the weighted CNN fits, which
were 0.70 and 0.76, depending on the weighting parameter.  Again, the predicted $\mu$ is
not sensitive to $s_{\sigma}$.  A more proper comparison of the prediction would be
to the unweighted CNN result, which gives $\mu = 0.61$, significantly below the predicted value.
Including an additional variance term with $\sigma_s = 0.25$ brings the predicted value down to 0.63,
much closer to the observed value.
Thus, while this method of predicting the modulation factor may be too high by 10-15\%,
adding an additional variance term may be sufficient to provide an accurate model of
the modulation as a function of the track angle uncertainty.

In reality, neither the Gaussian model nor the
(monopolar) von Mises model on $\cos (\psi - \psi')$ are likely to
be accurate descriptions of the angle error distributions.
Both models are appropriate only when the X-ray event interaction point can be reliably
identified by the NN.  Such will be the case for most long tracks but will not be valid for
short tracks (i.e., those with low energies) whose tracks have
approximately symmetrically distributed charge.
In this case, a dipolar von Mises model on $\cos 2(\psi - \psi')$ is closer to the truth.
For this model

\begin{equation}
\mu'(\kappa) = \frac{I_1(\kappa)}{I_0(\kappa)}
\label{eq:vonmises2}
\end{equation}

\noindent
\citep[e.g.][]{2021arXiv210708289P}.
However, this model cannot hold for all events because high energy tracks to not have equal
probability for $\psi$ as $\psi + \pi$.  Therefore, an alternative angle error distribution might be

\begin{equation}
G(\psi; \psi', \kappa) = \frac{1}{2 \pi I_0(\kappa) } [ f(\kappa) e^{\kappa \cos(\psi - \psi')} + \{1-f(\kappa)\} e^{\kappa \cos 2(\psi - \psi')} ]
\label{eq:double_vonmises}
\end{equation}

\noindent
where $f(\kappa)$ is a smooth function of $\kappa$, such that $0 \le f(\kappa) \le 1$.  For example, taking $f(\kappa) = \kappa/(1+\kappa)$
gives greater emphasis to the monopolar model for long tracks where $\kappa$ is large.
For $G(\psi;\psi',\kappa)$ as defined by Eq.~\ref{eq:double_vonmises}, the expected modulation factor is

\begin{equation}
\mu'(\kappa) = \frac{f(\kappa) I_2(\kappa) + [1-f(\kappa)] I_1(\kappa)}{I_0(\kappa)}
\label{eq:double_vonmises_mu}
\end{equation}

\noindent
Determining $f(\kappa)$ is outside the scope of this paper, requiring detailed CNN simulations.

At this point, we assume that a suitable functional form for $G(\psi;\psi',\alpha)$ can be
determined and validated with simulations and lab data. We now proceed to
show how the maximum likelihood formalism can be used to incorporate track angle
uncertainties.

\section{Revising the Likelihood}

Each event now has three important measurable
quantities associated with it that relate to the source polarization: energy $E$, phase angle $\psi$, and (generically)
angle uncertainty $\alpha$.  The event density in this 3D space is

\begin{equation}
\lambda(E, \psi, \alpha) = \int ~[ 1 +  (q \cos 2\psi' + u \sin 2\psi') ] G(\psi;\psi',\alpha) n_E A_E T P(\alpha;E) ~d\psi' 
\end{equation}

\noindent
where we now include $P(\alpha; E)$ to represent the probability density that an event of energy $E$ has
a track angle uncertainty $\alpha$.  This probability is normalized so that the integral over $\alpha$ is unity.
The log-likelihood of Eq.~\ref{eq:unbin} is rewritten to be

\begin{eqnarray}
S(n_0, q, u) & = & -2 \sum_i \ln \lambda(E_i, \psi_i, \alpha_i) +
2 T  \int n_E A_E dE \int P(\alpha; E) d\alpha  \int_0^{2\pi} [ 1 +  \mu'(\alpha)  (q \cos 2\psi + u \sin 2\psi) ]  d\psi \\
 & = & -2 \sum_i  \ln ( 1 +  q \mu_i \cos 2\psi_i + u \mu_i \sin 2\psi_i) - 2 N \ln(n_0) +
  4 \pi T n_0 A \int \mathcal{N}(E) \mathcal{A}(E) dE
\label{eq:new}
\end{eqnarray}

\noindent
which looks the same as before because $P(\alpha;E)$ is normalized and doesn't depend
on the source flux or polarization.
However, now $\mu_i = \mu'(\alpha_i)$ according to the appropriate choice
of error distribution (and any additional variance term) instead of $\mu_i = \mu(E_i)$.
At this point, the best estimates of $q$ and $u$ are the same as in Paper I:

\begin{eqnarray}
\hat{q} \approx \frac{\sum c_i}{ \sum c_i^2}\\
\hat{u} \approx \frac{\sum s_i}{ \sum s_i^2}
\end{eqnarray}

\noindent
and

\begin{eqnarray}
\sigma_q^2 \approx \frac{2}{\frac{\partial^2 S}{\partial q^2}} \approx \frac{1}{\sum c_i^2}\\
\sigma_u^2 \approx \frac{2}{\frac{\partial^2 S}{\partial u^2}} \approx \frac{1}{\sum s_i^2}
\end{eqnarray}

where $c_i \equiv \mu_i \cos 2 \psi_i$,  $s_i \equiv \mu_i \sin 2 \psi_i$, and the approximations hold for large $N$.
Note how these equations are similar to comparable equations derived by
\cite{2015APh....68...45K} and \cite{2018Galax...6...46V}.
Finally, the MDP is

\begin{equation}
\label{eq:mdp4}
{\rm MDP}_{99} \approx \frac{4.292}{(\bar{\mu^2} N)^{1/2}} ~~~~.
\end{equation}

\noindent
as in Paper I.

This analysis was based on an unbinned data set, where each event is tagged with uncertainty $\alpha_i$ and position
angle $\psi_i$.  Instead, if we suppose that the data are binned in small intervals $\Delta\alpha$ about $\alpha_j$,
with $C_j$ observed counts in bin $j$,
then the Eq.~\ref{eq:mdp4} would instead be written as

\begin{eqnarray}
\label{eq:binnedmdp} {\rm MDP}_{99} & \approx & \frac{4.292}{\sqrt{\sum_j \mu'(\alpha_j)^2 C_j}}\\
& = & \frac{4.292}{\sqrt{\sum_j e^{-4\sigma_j^2} C_j}} \label{eq:mdp_gaussian} \\
& = & \frac{4.292}{\sqrt{\sum_j (\frac{I_2(\kappa_j)}{I_0(\kappa_j)})^2 C_j}} \label{eq:mdp_vonmises} 
\end{eqnarray}

\noindent
similar to Eq. 50 of Paper I, except that $j$ represents an {\em uncertainty} bin, not an {\em energy} bin.
Eqs.~\ref{eq:mdp_gaussian} and \ref{eq:mdp_vonmises} apply to uncertainties given by Gaussian
and monopolar von Mises distributions, respectively.

While the derivation of the best fit polarization parameters does not depends on
applying weights to the likelihood, the computation of the MDP does depend on the spectrum of
the source via the $C_j$ terms.  These terms are

\begin{equation}
C_j = \int \lambda(E, \psi, \alpha_j) dE d\psi = N \eta_j ~~~~,
\label{eq:cj}
\end{equation}

\noindent
where the best estimate of $n_0$ from Eq.~\ref{eq:nzero} is substituted, and

\begin{equation}
\eta_j \equiv \frac{\int \mathcal{N}(E) \mathcal{A}(E) P(\alpha_j; E) dE}{\int \mathcal{N}(E) \mathcal{A}(E) dE } 
\label{eq:etaj}
\end{equation}

\noindent
gives the fraction of events in uncertainty bin $j$.
Thus, determining the MDP is merely a process of setting the number of counts $N$ in an observation and the assumed spectral
shape $\mathcal{N}(E)$.
Furthermore, once $N$ (or $n_0$) is set, the likelihood for $q$ and $u$ no longer depends
directly on the source
spectrum, so maximizing the likelihood over a set of CNNs only depends
on the values of $\alpha_i$ (via $\mu_i$) that are derived for the set of events.

\section{Summary}

We have outlined a way to to use the uncertainties in track angle measurements for an instrument
such as IXPE.  These uncertainties can be obtained by a machine learning method such as the
version based on convolutional neural networks by \citet{PEIRSON2021164740}.
The uncertainties can be directly related to the instrument modulation function, which is
critical to any analysis.  Furthermore, the uncertainties can be incorporated in a maximum
likelihood method for estimating polarization parameters by computing the distribution of
these uncertainties for the data set of interest.

All events are used in the analysis to determine the best fit parameters, obviating the need
for estimating an ``effective'' number of events.  Events with extremely large uncertainties would
have negligible modulation factors and would contribute very little to the estimation of $q$ and $u$.
Eqs.~\ref{eq:mdp_gaussian} through \ref{eq:etaj} show that the MDP depends primarily on
events with small uncertainties but all events are usable.

One caveat to this approach is that the angle error distribution should be properly
characterized.  We have demonstrated how the analysis may proceed for a Gaussian
distribution of errors, as provided by the current CNN method, but a von Mises
distribution is likely to be more appropriate.  Fortunately, the method outlined here
can accommodate whatever distribution is found to fit the data.
It is important to carry out enough simulations to estimate the $G(\psi;\psi',\alpha)$
(or just $\mu'[\alpha])$ and $P(\alpha; E)$ distributions.

\acknowledgments
Funding for this work was provided in part by contract 80MSFC17C0012 from the
Marshall Space Flight Center (MSFC) to MIT in support of IXPE, a NASA Astrophysics Small Explorers mission.
Support for this work was also provided in part by the National Aeronautics and
Space Administration (NASA) through the Smithsonian Astrophysical Observatory (SAO)
contract SV3-73016 to MIT for support of the Chandra X-Ray Center (CXC),
which is operated by SAO for and on behalf of NASA under contract NAS8-03060.
I thank Lawrence Peirson for reminding me of the von Mises distribution.

\bibliographystyle{aasjournal}                       

\end{document}